\documentclass[
    ,final          % use final for the camera ready runs
  ]
  {aipproc}

\layoutstyle{6x9}

\begin{document}

\title{Ferroelectric Dynamics in the Perovskite Relaxor PMN}

\author{P. M. Gehring}{ address={NIST Center for Neutron Research,
    National Institute of Standards and Technology, Gaithersburg, MD
    20899-8562} }

\author{S. Wakimoto}{ address={Department of Physics, University of
    Toronto, Toronto, Ontario, Canada M5S 1A7} }

\author{Z.-G. Ye}{ address={Department of Chemistry, Simon Fraser
    University, Burnaby, British Columbia, Canada V5A 1S6} }

\author{G. Shirane}{ address={Department of Physics, Brookhaven
    National Laboratory, Upton, NY 11973-5000} }

\begin{abstract}
  We review results obtained from recent neutron scattering studies of
  the lead-oxide class of perovskite relaxors PMN and PZN.  A
  ferroelectric soft mode has been identified in PMN at 1100~K that
  becomes overdamped near 620~K.  This is the same temperature at
  which polar nanoregions (PNR) begin to form, denoted by $T_d$, and
  suggests that a direct connection exists between the soft mode and
  the PNR.  The appearance of diffuse scattering intensity at $T_d$
  reported by Naberezhnov {\it et al.} lends further support to this
  picture.  At lower temperature the soft mode in PMN reappears close
  to $T_c = 213$~K (defined only for $E > E_c$).  These results are
  provocative because the dynamics below $T_c$ are characteristic of
  an ordered ferroelectric state, yet they occur in a system that
  remains cubic on average at all temperatures.  We discuss a
  coupled-mode model that successfully describes these data as well as
  those from earlier lattice dynamical studies of other perovskites
  such as BaTiO$_3$.
\end{abstract}

\maketitle

%%%%%%%%%%%%%%%%%%%%%%%%%%%%%%%%%%%%%%%%%%%%
%% MAINMATTER
%%%%%%%%%%%%%%%%%%%%%%%%%%%%%%%%%%%%%%%%%%%%

\section{Introduction - Soft Modes and PNR}

Recent neutron scattering studies of the perovskite ($AB$O$_3$)
relaxors Pb(Mg$_{1/3}$Nb$_{2/3}$)O$_3$ (PMN) and
Pb(Zn$_{1/3}$Nb$_{2/3}$)O$_3$ (PZN) have focused on the low-energy
($\le 20$~meV) lattice dynamics to determine whether or not a soft
mode picture is relevant to these chemically-disordered
systems.~\cite{pmg_prl1,pmg_prb,pmg_prl2} The answer to this question
is not obvious in spite of the similarities between PMN and PZN, and
the displacive ferroelectric PbTiO$_3$, which exhibits a phase
transition to a tetragonal structure at $T_c = 763$~K.  The occupation
of the $B$-site by either Mg$^{2+}$ or Zn$^{2+}$, and Nb$^{5+}$
cations creates a local charge imbalance, and gives rise to rapidly
varying random fields.~\cite{ye} Therefore anomalies in the lattice
dynamics of both PMN and PZN are to be expected.  Of particular
interest is the role played by the polar nanoregions (PNR), which for
PMN form between 600 and 650~K.~\cite{burns} Given their polar nature
they ought to couple strongly to polar phonon modes, including any
soft modes if present, in the relaxors PMN and PZN.

The existence of a soft transverse optic (TO) phonon mode has been
documented in numerous perovskites including PbTiO$_3$, BaTiO$_3$,
SrTiO$_3$, and KTaO$_3$.  In the case of PbTiO$_3$, the frequency of
the lowest-lying TO mode drops when cooled from high temperature and
condenses at the (first-order) transition temperature $T_c = 763$~K,
transforming the system into a tetragonal state.~\cite{shirane} Unlike
those in other perovskites, the soft mode in PbTiO$_3$ is well defined
except for a limited range of reduced wave vectors $q \le 0.08$.  The
zone center $q=0$ mode is, however, rather broad and well defined only
at temperatures far above $T_c$.~\cite{shirane}

Pioneering lattice dynamical studies were carried out by Naberezhnov
{\it et al.} on PMN from 300~K to 900~K.~\cite{naberezhnov} Fig.~1
shows the TO and TA phonon dispersions for PMN measured at 800~K
around (221).  A well defined zone center TO mode is observed at this
temperature.  However, it was identified as a hard TO1 mode in this
study because its structure factor was inconsistent with those
obtained from neutron diffuse scattering measurements by Vakhrushev
{\it et al.}~\cite{vakhrushev95} Prior Raman scattering studies had
also found no evidence of a soft mode in PMN.~\cite{Siny}

\begin{figure}
  \includegraphics[height=.5\textheight]{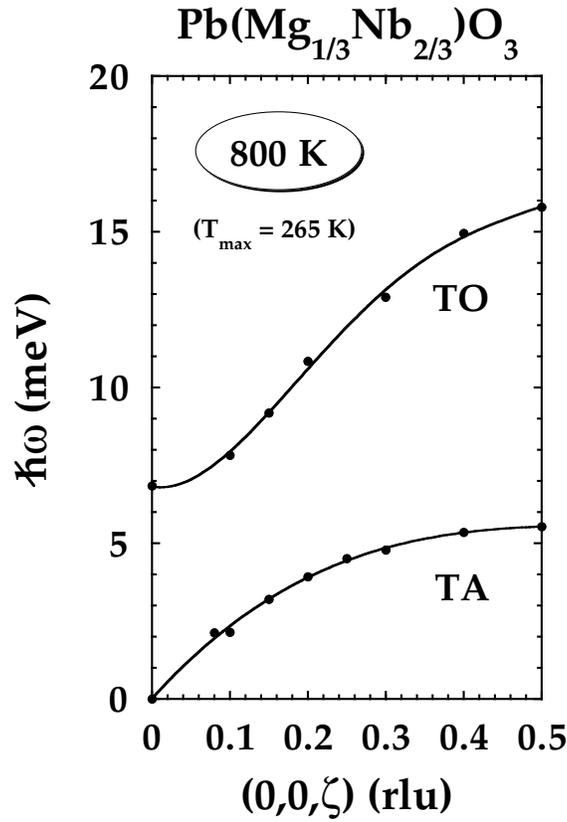}
  \caption{PMN dispersion measured at 800~K near (221).  Data taken
    from Naberezhnov {\it et al.}~\protect\cite{naberezhnov}}
\end{figure}

Subsequent experiments by Gehring {\it et al.} were done on PZN and
PZN doped with 8\% PbTiO$_3$ (PZN-8\%PT) for which large good quality
single crystals were readily available.  Fig.~2 combines a series of
constant-$\vec{Q}$ spectra measured on a 4.2~gm single crystal of PZN
at 500~K, nearly 100~K above $T_c$, into a greyscale contour plot (see
reference for experiment details).~\cite{pmg_prb} These data show an
unusual feature in which the TO branch {\it appears} to drop into the
TA branch around a $q$ of just under 0.15 reciprocal lattice units (1
rlu = 1.545~\AA$^{-1}$), which is a measure of the average size of the
PNR.  This anomaly, termed ``the waterfall'' because of its vertical
appearance, occurs as a result of a $q$-dependent damping in which the
TO phonon mode becomes increasingly damped with decreasing wave vector
$q$, i.\ e.\ increasing wavelength.  No phonon peaks are observed in
constant-$\vec{Q}$ scans for $q$ near the zone center, which indicates
that these TO modes are overdamped.  Thus the waterfall feature is
{\it not} a true dispersion because the scattering does not come from
propagating modes.  The waterfall is seen in PMN as well as in solid
solutions of both PZN and PMN and PbTiO$_3$ (PZN-$x$PT and PMN-$x$PT),
suggesting it may be common to all relaxor systems.~\cite{koo,gop}

\begin{figure}
  \includegraphics[height=.5\textheight]{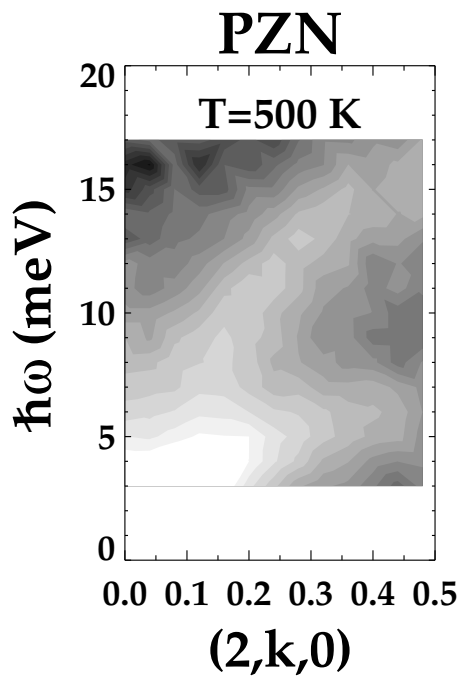}
  \caption{Contour plot derived from neutron inelastic 
    constant-$\vec{Q}$ scans measured on PZN at 500~K near (200).
    From Gehring {\it et al.}~\protect\cite{pmg_prb} Lighter regions
    indicate higher intensity.}
\end{figure}

In comparison to PbTiO$_3$, these results on PZN reveal the presence
of a new damping mechanism that effectively freezes out the
longest-wavelength TO modes and produces the unusual waterfall
feature.  The PNR are the most plausible cause of this damping.  To
test this idea would require heating the PZN crystal to temperatures
well in excess of $T_d \approx 760$~K where the PNR are no longer
present.  Because this was deemed too close to the decomposition
temperature of PZN, this experiment was performed on PMN, for which
$T_d \approx 620$~K as first determined by the optical measurements of
Burns and Dacol.~\cite{burns} As was the case in PZN, an overdamped
response was observed in PMN at $q=0$ at 500~K.  However the data in
Fig.~3 show unambiguous evidence of a well defined peak at $q=0$ at
1100~K that softens upon cooling.  At the same time the mode becomes
increasingly damped (broader energy linewidth).  Within experimental
uncertainty the soft mode becomes overdamped at $T_d$, thereby
establishing a definite connection between it and the PNR.  Equally
interesting is the fact that the square of the zone center TO mode
energy varies linearly with temperature ($(\hbar \omega_0^2) \sim (T -
T_0)$) as shown in the inset.  This is the hallmark of a ferroelectric
soft mode.  While the softening of this mode is clear, the
uncertainties in the data obtained from this small 0.1~cm$^3$ sample
do not permit a reliable determination of $T_0$, which should be close
to 400~K.~\cite{viehland} Data obtained on a much larger crystal
(0.4~cm$^3$) are discussed in a later section and are combined with the
data shown in Fig.~3 (see Fig.~5).

\begin{figure}
  \includegraphics[height=.5\textheight]{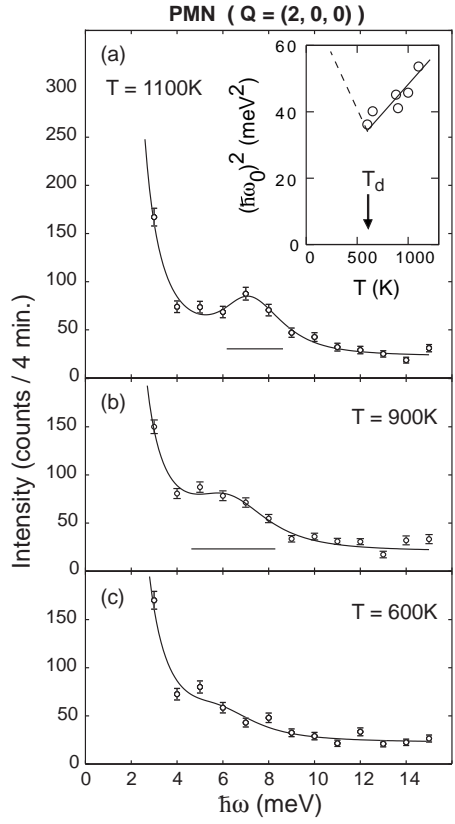}
  \caption{Temperature dependence of the soft TO mode in PMN.  From
    Gehring {\it et al.}~\protect\cite{pmg_prl2}}
\end{figure}

\section{The Connection to the PNR}

If the PNR are in fact the result of the condensation of the soft TO
mode in PMN, then the ionic displacements that characterize the
diffuse scattering intensities below $T_d$ must be consistent with
those that characterize the soft mode above $T_d$.  In particular,
since the ionic motions associated with the zone center TO vibrational
mode cannot shift the center of mass of the PMN unit cell, the static
displacements observed below $T_d$ must also satisfy this constraint.
But they do not.  Vakhrushev {\it et al.} measured the neutron diffuse
scattering intensities near 16 reciprocal lattice points on a single
crystal of PMN, and obtained the following relative ionic displacments
$\delta$ (normalized such that $\delta_{\rm Pb} = 1$): $\delta_{\rm
  MN} = 0.18$, $\delta_{\rm O(1)} = -0.738$, $\delta_{\rm O(2)} =
-0.549$.~\cite{vakhrushev95} These displacements do not satisfy the
center-of-mass constraint.  Yet if one believes that the static and
dynamic ionic displacements are related to the same polar state, then
one is faced with an unavoidable discrepancy that must be resolved.

\begin{figure}
  \includegraphics[height=.5\textheight]{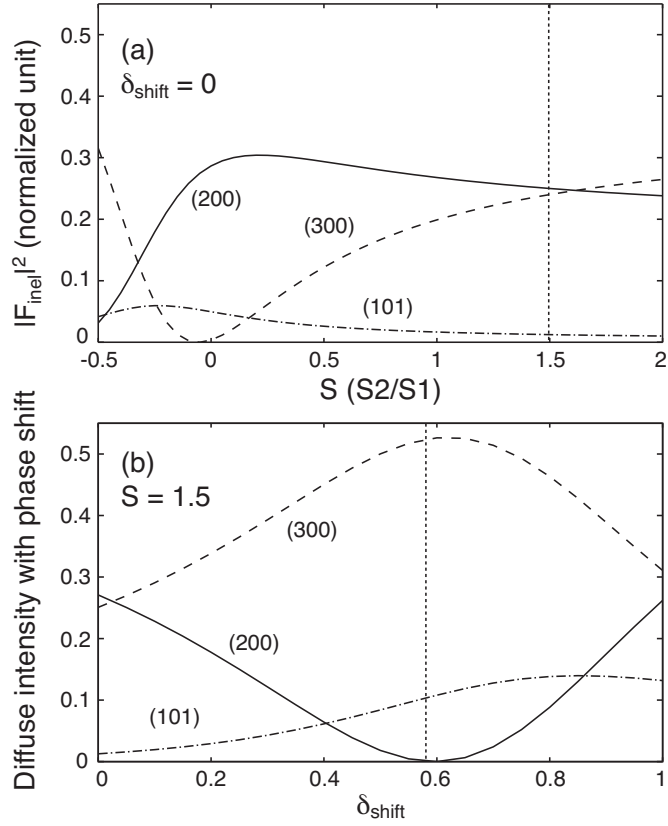}
  \caption{Model calculations of the inelastic and diffuse structure
    factors of PMN.  From Hirota {\it et al.}~\protect\cite{hirota}}
\end{figure}

This apparent contradiction was recently resolved by Hirota {\it et
  al.} who noticed that if a uniform shift of the proper magnitude
were subtracted from each of the ionic displacments, then the
remaining displacements satisfied the center-of-mass constraint.  In
other words, for each ion they assumed $\delta_i = \delta^{\rm cm}_i +
\delta_{\rm shift}$, where $i$ is the ionic index.  Using this model,
the size of the uniform shift was determined to be $\delta_{\rm shift}
= 0.58$, assuming an average oxygen displacement of $\delta_{\rm O} =
-0.64$.  The corresponding values for the $\delta^{\rm cm}_i$ are:
$\delta^{\rm cm}_{\rm Pb} = 0.42$, $\delta^{\rm cm}_{\rm MN} = -0.40$,
$\delta^{\rm cm}_{\rm O} = -1.22$.~\cite{hirota}

Calculations of both the inelastic and diffuse scattering structure
factors were made using these values.  Fig.~4 (a) shows the calculated
intensities of the TO phonon mode at the (101), (200), and (300)
reciprocal lattice points as a function of the ratio $S = S_2/S_1$ of
the Slater ($S_1$) and Last ($S_2$) modes.~\cite{harada} Inelastic
scans measured at (200) and (300) (see Fig.~6) fix the value of $S$
close to 1.5, consistent with that for PbTiO$_3$.  Fig.~4 (b) shows
the corresponding diffuse scattering intensities for this value of $S$
as a function of the single parameter $\delta_{\rm shift}$.  Agreement
with experiment, which indicates extremely little diffuse scattering
around (200), is obtained precisely at $\delta_{\rm shift} = 0.58$.
Because of the uniform shift $\delta_{\rm shift}$, the PNR are said to
result from a ``phase-shifted condensed soft mode.''  The shift of the
PNR is along their respective polar directions.

\section{Ferroelectric Dynamics in PMN below $T_c$}

Above $T_d \approx 620$~K PMN exhibits dynamical properties
characteristic of a normal ferroelectric in its cubic phase that are
surprisingly similar to those observed in PbTiO$_3$.  The formation of
PNR below $T_d$ appear to stifle the propagation of long-wavelength
polar TO modes due to their finite size.  The concurrent development
of diffuse scattering intensity at $T_d$, first observed by
Naberezhnov {\it et al.}~\cite{naberezhnov}, is consistent with the
picture of PNR condensing from the soft TO mode.  Yet PMN is reported
to remain cubic on average at all temperatures (at least down to
5~K).~\cite{mathan} On the other hand, it is known that a macroscopic
ferroelectric phase can be induced in PMN by cooling it in a small
electric field.  After removal of this field, the ferroelectric state
in PMN is lost via a first-order transition when heated above $T_c =
213$~K.~\cite{ye2,ye3} Therefore the lattice dynamical studies of PMN
were extended to low temperature, placing special emphasis on the
temperature region near $T_c$.

\begin{figure}
  \includegraphics[height=.5\textheight]{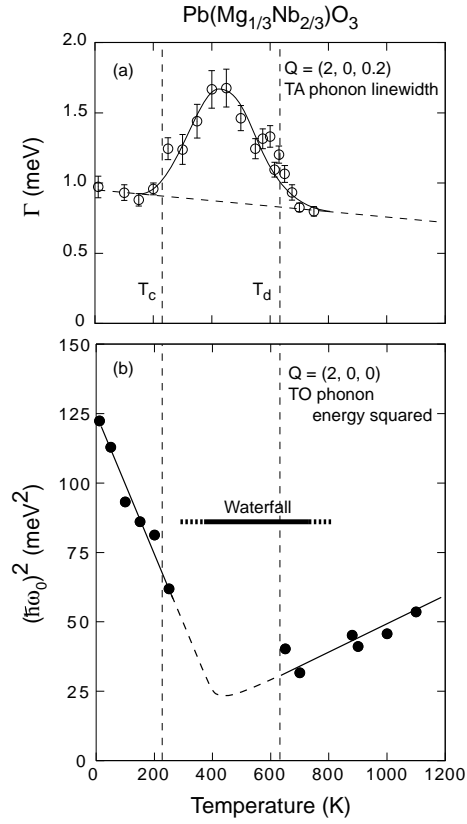}
  \caption{Summary of the TA linewidth and TO soft mode temperature
    dependence in PMN.  From Wakimoto {\it et
      al.}~\protect\cite{wakimoto}}
\end{figure}

Fig.~5 summarizes recent neutron inelastic scattering results obtained
by Wakimoto {\it et al.} on a large 0.4~cm$^3$ single crystal between
10 and 750~K.~\cite{wakimoto} By far the most remarkable feature is
the recovery and subsequent behavior of the soft mode where $(\hbar
\omega_0)^2 \sim (T_0 - T)$ below $T_c$ shown in Fig.~5 (b) that is
typical of many ferroelectrics in their ordered states.  This implies
that a phase transition from the relaxor state to a low-temperature,
short-range ordered ferroelectric state, occurs at $T_c$ in the
absence of an applied field.  But this contradicts the average cubic
symmetry of PMN.  However, if the length scale of the ferroelectric
order were $\sim 100$~\AA, then one could plausibly resolve this
contradiction since this length scale would appear as long-range order
in neutron phonon measurements, but short-range order to x-rays.
Indeed, regions of ferroelectric order $\sim 100$~\AA\ in size have
been reported by de Mathan {\it et al}.~\cite{mathan} Hypersonic
damping \cite{Tu} and Raman scattering \cite{Siny} studies have also
reported anomalies at or near $T_c$, as has a change in the topology
of the PNR,~\cite{Vakhrushev}, all of which lend support to the idea
of a low-temperature ferroelectric state in PMN.  It is interesting to
note that in a zone center mode is observed at low (25~K) temperature
in PZN at essentially the same energy as that found in PMN
(10.5~meV).~\cite{pmg_prb}

Fig.~5 (a) shows an anomalous damping of the TA mode that coincides
with the overdamping of the soft mode and the appearance of the PNR at
$T_d$.  This suggests a non-uniform distortion of the lattice which
then vanishes below $T_c$, ostensibly when the short-range ordered
ferroelectric phase is established.  A similar TA broadening has also
been observed in PMN-20\%PT by Koo {\it et al.}~\cite{koo}

\section{Mode Coupling Analysis}

The scattering cross section used to fit these inelastic data is based
on the coupled-mode description first used by Harada {\it et al.} to
fit asymmetric phonon lineshapes observed in BaTiO$_3$.~\cite{harada2}
Typical data measured in two different zones, (200) and (300), and
taken above $T_d$ at 690~K, are shown for PMN in Fig.~6.  The
coupled-mode cross section is parametrized by a coupling-constant
$\lambda$, along with inelastic structure factors $F_1$ and $F_2$ for
the coupled TO and TA modes, respectively.  Additional parameters are
the phonon energies $\hbar \omega_1$ and $\hbar \omega_2$, and the
linewidths $\Gamma_1$ and $\Gamma_2$.  All of these parameters were
allowed to float freely between the two scans.  The values for the TA
and TO phonon energies were identical to within 0.1~meV between (200)
and (300), as required, with $\hbar \omega_1 = 10.8$~meV, $\hbar
\omega_2 = 4.2$~meV, and represent a useful cross check of the
integrity of the data.  However the ratio of the dynamic structure
factors $F_1/F_2$ is markedly different in each zone.  At (200) the
ratio is 2.3, whereas at (300) it is 27.5, nearly an order of
magnitude different.  The coupling constant $\lambda$ has nearly the
same absolute value in each zone (-14.1 versus 17.6 for (200) and
(300) respectively), but opposite sign.  Physically this is due to a
sign change in the TO structure factor $F_1$, not $\lambda$.

\begin{figure}
  \includegraphics[height=.5\textheight]{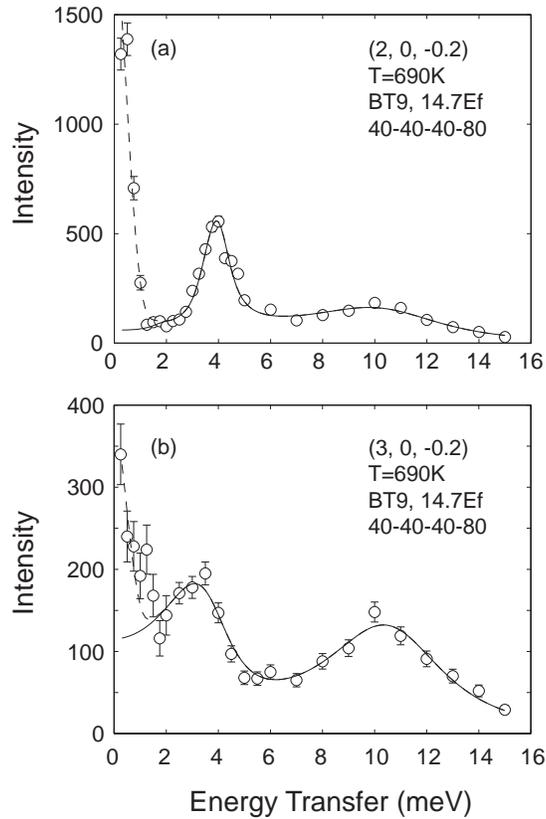}
  \caption{Constant-$\vec{Q}$ scans measured in PMN around the (a) (200)
    and (b) (300) Brillouin zones.  Solid lines are fits to a coupled
    TA-TO mode cross section, corrected for instrumental resolution
    effects (data taken from Hirota {\it et
      al.}).~\protect\cite{hirota} }
\end{figure}

Fig.~7 conveys the manner in which the coupled-mode cross section
varies with the TO mode linewidth $\Gamma_1$ (corresponding to an
increased damping), as well as with a change in $\lambda$.  This cross
section gives an excellent model description, shown in panel (a), of
the $q$-dependent damping observed in both PZN and PMN below $T_d$.
Two cases are presented in panels (b) and (c) following Bullock {\it
  et al.}~\cite{bullock} Panel (b) corresponds to well-separated TA
and TO modes with energies of 5 and 10~meV, respectively, and
identical linewidths of 1~meV.  Panel (c) corresponds to TA and TO
modes with energies of 6 and 8~meV, and substantially different
linewidths of 0.5 and 4~meV, respectively, such that there is
substantial spectral overlap.~\cite{bullock} In both cases there is a
clear transfer of spectral weight from the TO mode to the TA mode when
the mode coupling is turned on ($\lambda \ne 0$).

\begin{figure}
  \includegraphics[width=0.9\textwidth]{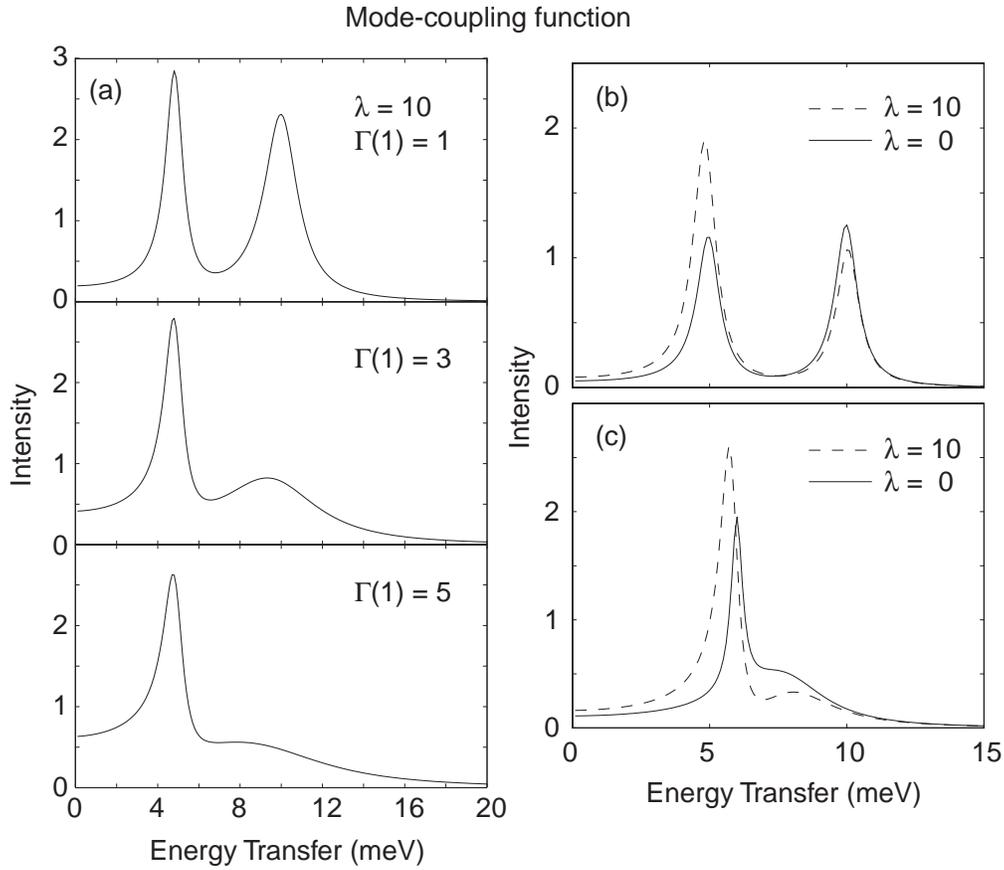}
  \caption{(a) Variation of coupled-mode cross section with TO linewidth
    $\Gamma_1$. Comparison of the coupled-mode cross section with and
    without coupling when modes (b) are well-separated, and (c) have
    significant spectral overlap.  Parameters used are identical to
    those in Fig.~7 by Bullock {\it et al.}~\protect\cite{bullock}}
\end{figure}

\section{Conclusions}

We have given a brief overview of the lattice dynamics of PMN and PZN.
In PMN a well-defined soft mode is observed at 1100~K that becomes
increasingly damped upon cooling, and eventually becomes overdamped at
$T_d$.  The overdamping of the soft mode and the appearance of the
diffuse scattering both at $T_d$ strongly point to a direct connection
between the soft mode and the PNR.  The resolution of the discrepancy
between the static and dynamic structure factors using the
``phase-shifted condensed soft mode'' model of Hirota {\it et al.}
provides a convincing framework in which this connection can be
understood.  The damping of the soft mode that is present at
temperatures above $T_d$ is believed to be a result of the fact that
the PNR exist as dynamical entities at higher temperature.

The soft mode dynamics of PMN below $T_c$ are characteristic of an
ordered ferroelectric phase, and stand in contrast to the average
cubic structure reported for PMN.  The assumption of short-range order
($\sim 100$~\AA) can be used to reconcile these two sets of
experimental observations.  At all temperatures, a coupled-mode cross
section, properly convolved with the instrumental resolution function,
provides an excellent description of the neutron inelastic data.

A very different interpretation of the soft mode dynamics in PMN has
recently been proposed by Vakhrushev and Shapiro \cite{vs} in which
the existence of a soft ``quasi-optic'' (QO) mode, distinct from the
zone center TO mode, is claimed to explain the observed Curie-Weiss
behavior of the dielectric susceptibility above $T_d$.  The QO-mode is
derived from a mode-coupling analysis of neutron inelastic lineshapes
measured between 490 and 880~K, and is attributed to the intrinsic
chemical disorder of PMN.  More studies of PMN at higher resolution
are clearly needed to reconcile these different points of view.

\begin{theacknowledgments}
  We thank A.\ A.\ Bokov, K.\ Hirota, V.\ Kiryukhin, T.\ -Y.\ Koo, K.\ 
  Ohwada, S.\ M.\ Shapiro, C.\ Stock, N.\ Takesue, S.\ B.\ Vakhrushev,
  and H.\ You for stimulating discussions.  Financial support from the
  U.\ S.\ DOE under contract No.\ DE-AC02-98CH10886, and the Office of
  Naval Research under Grant No.\ N00014-99-1-0738, is acknowledged.
  Work at the University of Toronto is part of the Canadian Institute
  for Advanced Research and is supported by the Natural Science and
  Engineering Research Council of Canada.  We acknowledge the support
  of the NIST Center for Neutron Research, the U.\ S.\ Department of
  Commerce, for providing some of the neutron facilities used in the
  present work.
\end{theacknowledgments}

%% If the bibliography is produced without BibTeX comment out the

\bibliographystyle{aipproc}   % if natbib is available

%\bibliography{sample}

\end{document}